\documentclass[12pt,letterpaper]{iopart}

\usepackage{epsfig,graphics} 
\begin{document}
\title[Semi-relativistic approach to three body problem]{A semi-relativistic
  approach to the circular restricted three body problem 
  and a numerical experiment around the 3:2 resonance}
\author{Eduardo Gu\'eron}
\address{Dep. de Matem\'atica Aplicada-IMECC Unicamp, SP Brazil
  13083-970\footnote{Part of the work developed at EAPS-MIT, Massachusetts
  02139} }  
\ead{\mailto{gueron@ime.unicamp.br}}    
\begin{abstract}
A  Hamiltonian that approaches the study of the three-body problem
in general relativity is obtained. We use it to 
study the relativistic version of the circular restricted
three-body problem in which the first body is the heaviest and
the third body is a test-particle. We focus  on the orbits
around the 3:2 resonance. We show that, in spite of the notable difference
between the relativistic and Newtonian orbits, most of the resonant region is
preserved. Nevertheless, differently from the Newtonian case, 
the frequencies between the second and the third body are no longer
commensurable.  
\end{abstract}

\section{Introduction}

In the Newtonian context, the restricted three body problem has been 
widely studied in order to  predict resonances in the solar system dynamics, 
specially concerning the motion of
comets and asteroids (see for instance~\cite{Danby,solar,Henon}). 
By definition, the third body is a test particle
that does not gravitationally affect the motion of the other two bodies that
are in a Keplerian motion. A typical example is the study of the  motion of
the asteroids between Jupiter and the Sun, considered to be the first two
bodies. Most of these asteroids are in a mean-motion resonance with
Jupiter~\cite{lemaitre}. An interesting problem is the existence of gaps in
the distribution 
of semi-major axes of the orbits of asteroids - the Kirkwood gap. Wisdom
solved this problem by studying the orbits near the 3:1
commensurability~\cite{jw}. Regarding  the three body problem  we may also
cite the problem of the Hannay angle~\cite{hannay,spal1} and a new exact
solution obtained by Chenciner and Montgomery~\cite{annals}.

 Perhaps the first article
concerning the n-body problem in General Relativity was written by De Sitter
\cite{desitter} 
in the very beginning of the last century and his predictions about the
precession of the Moon was
confirmed by Shapiro {\it et al}~\cite{shapiro}. In a relativistic framework, 
approximations are always needed in order to study the
gravitational field generated by more than one body. Post-Newtonian
approaches  were employed several times in the three body
problem, one may 
cite the work from  Krefetz~\cite{krefetz} in which a solution corresponding to
the equilateral Lagrange one is obtained. Also, Rosswog and
Trautmann~\cite{rosswog} studied the Lagrangian 
stability points for the circular problem. Brumberg~\cite{brumberg} has
considered such problem in a
radiating binary system. Using a different approach, Gu\'eron
and Letelier~\cite{gl} considered the interaction among orbiting particles
around a black hole.  We also may cite the work from Campanelli {\it et
  al}~\cite{campanelli} where  the influence of a very massive black hole in
the stability of binary systems by means of Post-Newtonian techniques was
studied. 

In this article, I 
introduce a novel formalism to account for the General Relativistic effects on 
the n-body problem for systems so that there is a predominant very massive
body, for instance, a super massive black hole as the one in the center of the
Milky Way~\cite{ghez}.  The technique presented  is based
on the Hamiltonian formalism for the geodesic 
problem. The gravitational effects of the heaviest source  are associated to
an exact solution of the Einstein field equations and the field of the other
bodies  perturbs the corresponding Hamiltonian 
as a classical gravitational potential energy but does not affect the motion
of the heaviest body. Therefore equations of motion obtained describe the
trajectory of the light bodies. We can compare the method proposed here with
the ones used to compute relativistic secular effects in the orbits of solar
system objects~\cite{long}. In this procedure the mean
relativistic 
corrections are given continuously with the Hamiltonian, differently to other
approaches in which such effects are considered periodically in the correction
of the planetary orbits.  

As a first test to this formalism, the restricted and circular three body
problem is numerically studied. For this purpose, the Schwarzschild solution
is associated to the heaviest source (hereafter the first body). The computed
Hamiltonian is changed by the potential energy generated by a second body in a
circular geodesic motion around the first and the equations of motion of a
test-particle (third body) affected by these two bodies are derived. 
The stability of the orbits around the 3:2 central resonance is studied by
means of Poincar\'e sections.  This specific choice is justifiable since
chaotic regions are clearly distinguished from resonant islands in the
circular problem. It was found that the general relativistic
effects destroy the commensurability of the frequencies of the orbits
of the second and the third bodies mainly for eccentric orbits. 
Remarkably, most of the resonant regions are preserved. In other words, it
means that most of the orbits that were stable in the Newtonian limit remains
stable when relativistic effects are considered. Nonetheless the ratio between
the motion period of the second body and the period of the motion of the third
one is no longer a rational number\footnote{Numerically the number is considered``irrational'' if, in a reduced form  $p/q$, one has that the minimum $p,q
  \gg 1$.}.  

\section{Hamiltonian Formalism}

Given a metric tensor $g_{\mu\nu}$, the geodesic equations in General
Relativity are 
equivalent to 
the Euler-Lagrange ones obtained from the Lagrangian
\begin{equation}
{\mathcal L}=\sqrt{|g_{\mu\nu}u^{\mu}u^\nu|},\label{lag1}
\end{equation}
$u^\mu=dx^\mu/d\tau$ and the units are so that $c=1$ and $G=1$. The Einstein
sum convention is adopted and Greek and Latin indices vary respectively from
0 to 3 and 1 to 3. 

 Usually the
equations of motion are written with an 
affine parameter, i.e., a free variable proportional to the arc length (for
instance the proper time $\tau$).  Sometimes,
however, the use of a non-affine parameter becomes necessary. For
this purpose, one must derive the corresponding  formulation to the
geodesic problem. A typical example of a non-affine useful parameter is the
coordinate time. When the evolution of more than one particle is to be
compared, $t$ as a free parameter seems to be more adequate. 

In order to write the Hamiltonian for a test particle whose orbit is
parameterized by a time-like coordinate, one starts with
the action
\begin{equation}
S=\int p_\mu dx^\mu = \int\left( p_0+p_i\frac{dx^i}{dt} \right) dt,
\end{equation}
where $p_\mu=m g_{\mu\nu}\frac{dx^\nu}{d\tau}$. Recalling the Legendre
transformation $H=p_i\frac{dx^i}{dt}-{\mathcal L}$ one may identify  $H=-p_0$.

Writing  $p_0$ as a function of the other components (the relation
$g^{\mu\nu}p_\nu p_\mu=-m^2$ was considered)  one
gets the Hamiltonian:  
\begin{equation}
H(p_i,x^j,t)=\frac{g^{0i}}{g_{00}}+\left (
\frac{(g^{ij}p_ip_j+m^2)}{-g^{00}}+\left ( \frac{g^{0i}p_i}{g^{00}} \right )^2
\right ) ^{1/2}. \label{hamg0}
\end{equation}
This formulae is general and we only require $t$ to be a
time-like coordinate. For more details about this formulae see the
Refs.~\cite{wald,apostila}. 

Now let us suppose that besides the gravitational field generated by the
 central source, there are  smaller sources that affect the motion of the
 test particle.   The matter we have now is how to include such field in the
 Hamiltonian formalism. Assuming that these sources are much lighter than the
 main source, one can compute their contribution by adding a perturbative term
 in the above given Hamiltonian Eq.~(\ref{hamg0}). The natural choice is 
 the Newtonian gravitational potential.

Then I propose that for a  system with a heavy source (or a predominant mean
field) and $n$ smaller massive bodies, can be approximated by
the following Hamiltonian:
\begin{equation}
H(p_i,x^j,t)=\frac{g^{0i}}{g_{00}}+\left (
\frac{(g^{ij}p_ip_j+1)}{-g^{00}}+\left ( \frac{g^{0i}p_i}{g^{00}} \right )^2
\right ) ^{1/2}-\sum^n_{k=1}
\frac{\mu_k}{|\mathbf{r}_k-\mathbf{r}|},\label{hamg1} 
  \end{equation}
where $\mu_k$ is the mass of the $k^{th}$ body. Notice that the dependence on
the mass of the test particle was eliminated. 

\subsection{Limits of applicability} 

The approximation presented above is quite simple. The Newtonian potential is
added to the exact relativistic Hamiltonian that leads to the geodesic
equations parameterized by the coordinate time. Therefore it is not difficult
to enumerate the relativistic effects that are not being considered in this
approach. First one must have in mind that in the exact solution there should
be a  lot of ``mixed'' terms in the Hamiltonian due to the nonlinearity
of the Einstein equations - in the Newtonian approach they do not
exist. A didactic example is the solution of two static black holes sustained 
by a string or a strut~\cite{weyl,gueron}. The metric and hence the
Hamiltonian are much more complicated than a simple sum of potential energies.
Also, terms due to the non-staticity of the solution are not being
considered, the obvious neglected effect is the damping by gravitational
radiation~\cite{cooperstock,hulse,burko,quinn}.  

Although these terms are very important in some astrophysical systems (see for
instance~\cite{barack, mino, spallicci}), the
largest error  corresponds to the fact that  all the source but the
central one are considered Newtonian. We may say that the method can be used
when  $\sum_i^N (M/d_{0i})\gg \sum_{i<j}^N (m_i/d_{ij})$ that is the
relativistic effects due to 
the central source are  much more important than the relativistic
contributions from the lighter bodies ($d_{0i}$ is the $i^{th}$ particle
distance to the central body and 
$d_{ij}$  between the $i^{th}$ and the $j^{th}$ orbiting particles supposing
$m_i\ge m_j$).  Notice that, since the
first term of the function (\ref{hamg1}) is associated to a mean gravitational
field, an 
average of some relativistic effects due to the orbiting bodies might be
computed in the Hamiltonian. An example is the use of multipolar solution
\cite{Quevedo89,boisseau}.        

A simple idea for estimating the error starts with the Schwarzschild metric
$ds^2=(1-2M/r)dt^2-1/(1-2M/r)dr^2+r^2d\Omega^2$
(that will be better explored in the next sections). We could think that in
the full general relativistic solution, the perturbative term would be similar
to the first one. This, of course, is not true due to the nonlinearity of
the Einstein equations. Nevertheless, it may be used to estimate the order of
magnitude of the error we are doing. For the Schwarzschild metric above
written we have that 
\begin{equation}
[g^{\mu\nu}]={\rm diag} \left
[1/(1-2M/r),1-2M/r,(r\sin \theta)^{-2},r^{-2}\right ] \label{contra}
\end{equation}
 Now, if the motion
of the particle is restricted to the plane $\theta=\pi/2$ we will have that
$p_\theta=0$. With these assumptions, we might expand the 
expression corresponding to Hamiltonian of a single particle with planar
motion in the Schwarzschild line element substituting (\ref{contra}) in
(\ref{hamg0}): 
\begin{equation}
H_s=\sqrt{(1-2m/r)((1-2m/r)p_r^2+p_\varphi
    ^2/r^2+1)}.
\end{equation}
 The first relativistic term is of the order $m^2/r^2$ in
    geometric units. Roughly speaking this term has the order of magnitude of
    the error done in the simulations.

\subsection{Schwarzschild}

The particular system I am interested in this paper concerns planar motions
around 
a black hole (all the considered bodies are in the same plane). The
Hamiltonian for a free particle around a Schwarzschild black hole with mass
$M$  can be cast as:
\begin{equation}
H=\sqrt{(1-2M/r)((1-2M/r)p_r^2+p_\varphi
    ^2/r^2+1)},  
\label{hrg}
\end{equation}
in which usual spherical coordinates are used. 

The Hamilton equations obtained from the above function will give the
geodesic motion of a test particle  orbiting around a static and non charged
black hole. Since I want to
study  the motion of a third body that is attracted  by the central
source  and the orbiting body. It will be considered that the force exerted by
the second body on the 
third one is essentially Newtonian. For this achieving, one  assumes that the
central
source is fixed and the motion of the second body is given by the
time-like  geodesic around the first one
represented by the curve ${\bf r_2}(t)$.  Using these assumptions, we may
write a Hamiltonian  that approximates the motion of the third body by
adding the contribution of the second body as a Newtonian gravitational
potential energy to Eq.\ref{hrg}. Thus the proposed Hamiltonian has the form 

\begin{equation}
 H=\sqrt{(1-2M/r)((1-2M/r)p_r^2+p_\varphi
    ^2/r^2+1)}-\frac{m}{|{\bf r}-{\bf r_2}(t)|}. \label{ham} 
\end{equation}
(M and m are the masses of the first and second bodies respectively)

Summarizing, the first term of this Hamiltonian represents the geodesic motion
of a test-particle around a spherical symmetric source and the second  is
the gravitational potential generated by an orbiting body. Therefore, the
Hamilton-Jacobi equations lead to the equations of motion of the test-particle
(Remind that in this particular example all the three bodies are in the same
plane)

\section{Simulations}

I intend to make an analysis of the orbits by means of the study of surfaces
of   
sections in the phase space. Since the Hamiltonian is
time-dependent, it is useful to study periodic orbits of the second body
in order to eliminate this dependence.  Nevertheless, periodic orbits
are not easily found in time-like geodesics around a black hole except when
the test particle experiments a circular motion. Therefore it is  
interesting to study the restricted three body problem when the second
 orbit is circular, i.e., when the second body has the motion 
\begin{equation}
\varphi_2=\omega t,\:\:\: r_2 =\mbox{constant.}
\label{2mot}
\end{equation}
Substituting Eq.\ref{2mot} in Eq.\ref{ham} one gets the Hamiltonian that
describes the motion of the third body in this particular case
\begin{equation}
\!\!\!\!\!\!\!\!\!\!\!\!\!\!\!\!\!\!\!\!\!\!\!\!\!\!
    H=\sqrt{(1-2m/r)((1-2m/r)p_r^2+p_\varphi 
    ^2/r^2+1)}-\frac{m}{\sqrt{r^2+r_2^2-2rr_2\cos(\omega t-\varphi)}}.
    \label{hamc}  
\end{equation}
It is clear that  $H$ is not a constant of motion because of its time
dependence . However by means
of the simple 
conjugate transformation of coordinates~\cite{jw} $\varphi'=\omega t-\varphi$
(the other coordinates do not transform)
one writes  a new Hamiltonian $K$ that is constant
\begin{equation}
\!\!\!\!\!\!\!\!\!\!\!\!\!\!\!\!\!\!\!\!\!\!\!\!\!\!\!\!\!\!\!\!
    K=\sqrt{(1-2M/r)((1-2M/r)p_r^2+p_\varphi 
    ^2/r^2+1)}+\omega
    p_\varphi- \frac{m}{\sqrt{r^2+r_2^2-2rr_2\cos(\varphi')}} 
    \label{hamk}  
\end{equation}

 The constant angular speed of the second body $\omega$ may be calculated from
 Eq.\ref{hrg} and $g_{\mu\nu}u^{\mu}u^\nu=-1$. One gets  that
  $\omega=\frac{d\varphi_2}{dt}=\sqrt{\frac{m}{r_2^3}}$ - 
 exactly the same value obtained in the Newtonian analogue.

With a constant Hamiltonian the problem now has two degrees of freedom.
Therefore it is easy to make a study on the stability by means of  Poincar\'e
sections. For this purpose I mark the position where an
orbit crosses the plane defined by $\varphi-\omega t=\varphi'=2k\pi$, i.e.,
when the 
test-particle and the second body are in the same angular position ($k$ is an
integer).   

The first step is the determination of the constant $K$ that defines a
three dimensional hypersurface in the phase space inside which the motion must
be confined. For this purpose, I calculate the parameters of the circular
geodesic whose period is  2/3 of the period the second. This requirement gives
us $r$ and $p_\varphi$. Those values with the 
parameters of the second body are used to fix $K$ in the simulations. Note
that $K$ can be associated to 
the Jacobi's usually  computed in restricted three body problem
\cite{Danby,Henon}.    

After fixing $K$ as described above, I determine a set of initial conditions 
in order to plot the Poincar\'e section -  there are three free variables to
be chosen. Each figure to be presented is associated to a fixed $K$ and out of
them it was considered $m/M=10^{-3}$. For distances of the order of 10
Schwarzschild radii to the second body (units defined with respect to the
first body), the error will be of the order of $10^{-6}$.

The procedure adopted to plot the Poincar\'e section has an important
characteristic: If we had the Newtonian problem, all the Poincar\'e sections
would present exactly the same aspect. In other words they correspond to the
same Jacobi constant. When we change proportionally the initial
positions and velocities of the bodies without changing their masses, we obtain
qualitatively the same situation in the Newton gravitational problem but a
different situation in the relativistic problem. 

\begin{figure}[p] 
\includegraphics[width=13cm]{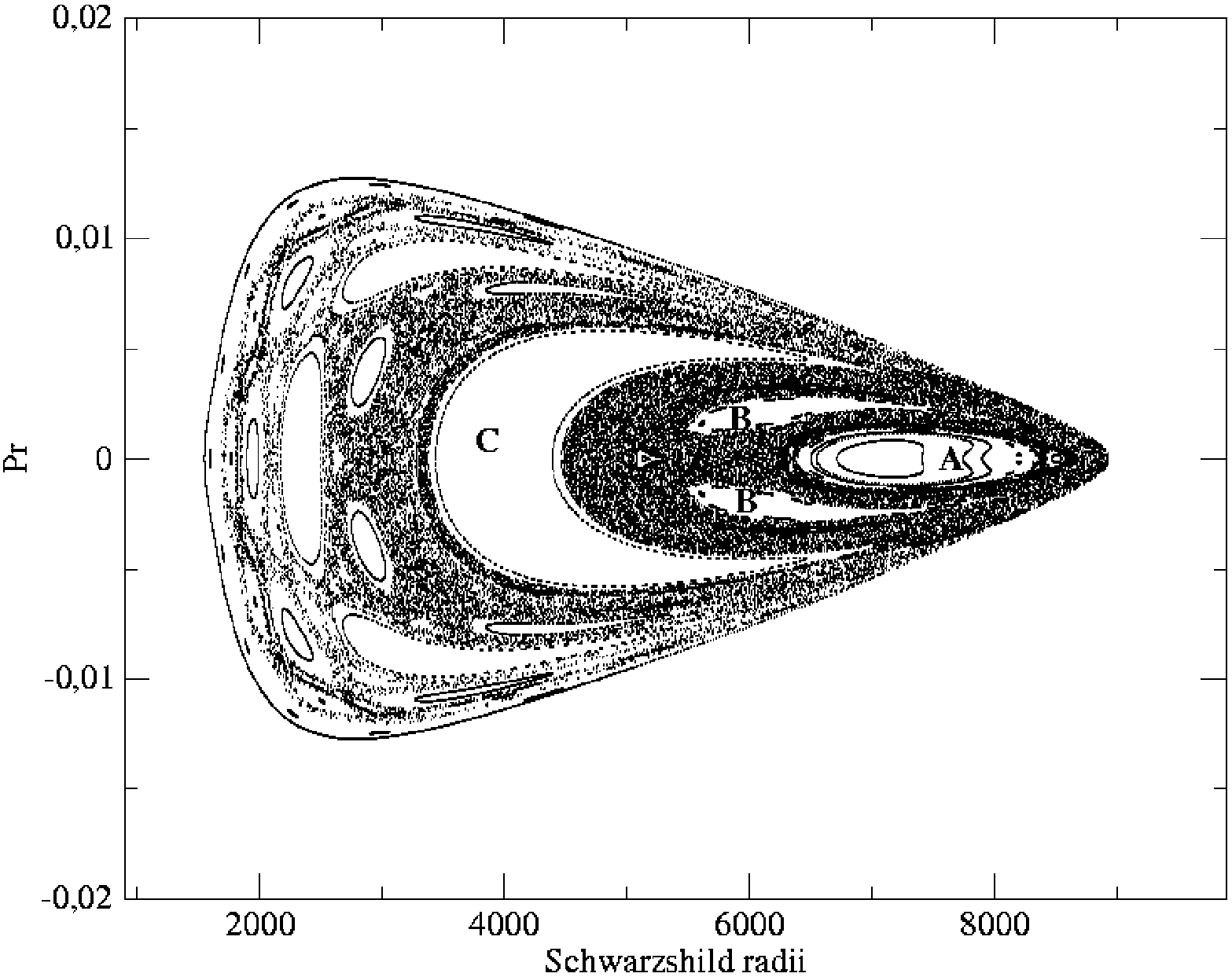} 
\caption{Surface of section distant to the black hole, $K= 0.9998$. The
 radius of the circular orbit $r_2=10^4$ (see Eq.\ref{hamk} ). The ordinate is
 the radial component of the momentum and in the abscissas one has the
 distance of the test-particle to the central mass.  In this
  example a situation where the relativistic effects do not play an important
  role. The resonant regions are very clear. For example, I mark {\bf A} -
  3:2, {\bf B} -  5:3 and {\bf C} - 2:1 resonances}   
\label{longe} 
\end{figure}

\begin{figure}[p] 
\includegraphics[width=13cm]{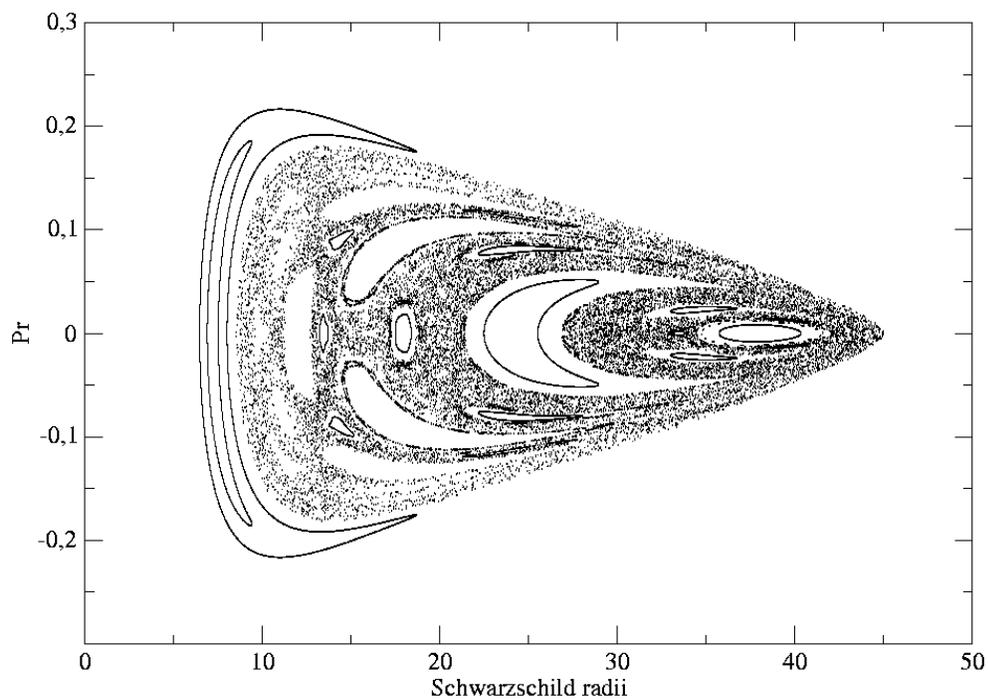}  
\caption{For $K=0.969$, the particles are much closer to the central
  source and  relativistic effects are very important, $r_2=50$. In spite of
  that, most of the  stable islands are preserved when compared to the
  previous figure.}    
\label{perto} 
\end{figure}

The Poincar\'e sections are then presented in Fig.\ref{longe}-\ref{perto}. In
all the figures, the abscissa is the radial coordinate (the distance of the
test particle to the central mass) and the ordinate is its conjugated
momentum. Both are computed when $\varphi-\omega t=\varphi'=2k\pi$ as
described above.

 In Fig.\ref{longe}, the relativistic
aspects do not play an important role  since the orbiting bodies are very
distant from the central source thus one has a
typical surface of section for the classical problem - notice that indirectly
the Newtonian limit of the Eq.\ref{hamc} is tested here. Some resonant regions
are presented and the eccentricity of the test-particle orbits increases as I
increase the rate $\langle d\varphi/dt\rangle/\omega$. The circular orbit are
inside the 
$3/2$ resonant island. I mark
the 3:2 resonant island {\bf A}, 5:3 with {\bf 
  B} and 2:1 with {\bf C}.(Along this paper only mean orbital resonances are
considered.)  These values are determined by computing
$\langle d\varphi/dt\rangle/\omega$,  it means that the period of motion of
the second body is 3/2 times the period of the test particle (third body) when
they do not interact. Since the points
of the surface section are marked when the orbits are in conjunction
with respect to the central source, if they are in a
$R:S$ resonance, the number of stable islands will be $|R-S|$.

Repeating the procedure we get the Fig.\ref{perto} in which the orbits are very
close to the central source and the 
relativistic effects are important. Nevertheless, most of the stable
islands are preserved. It is clear that the islands corresponding to the
3:2, 5:3 and 2:1 resonances are preserved when compared to the
Fig.\ref{longe}. (In this case the ratios between the periods in the resonant
islands are not necessarily commensurable as we will discuss later.) 

Another important aspect concerns  chaotic orbits. Close to the
central source  they tend to fill a larger portion of
the phase space comparing with the Newtonian case. In a
extreme case, i.e., very close to the central source or for huge masses of the
first 
body, the only bounded orbits are the ones that belong to some resonant
region, thus,  almost all the chaotic orbits become unbounded, see
Fig.\ref{escapa}.

\begin{figure}[p]
\includegraphics[width=13cm]{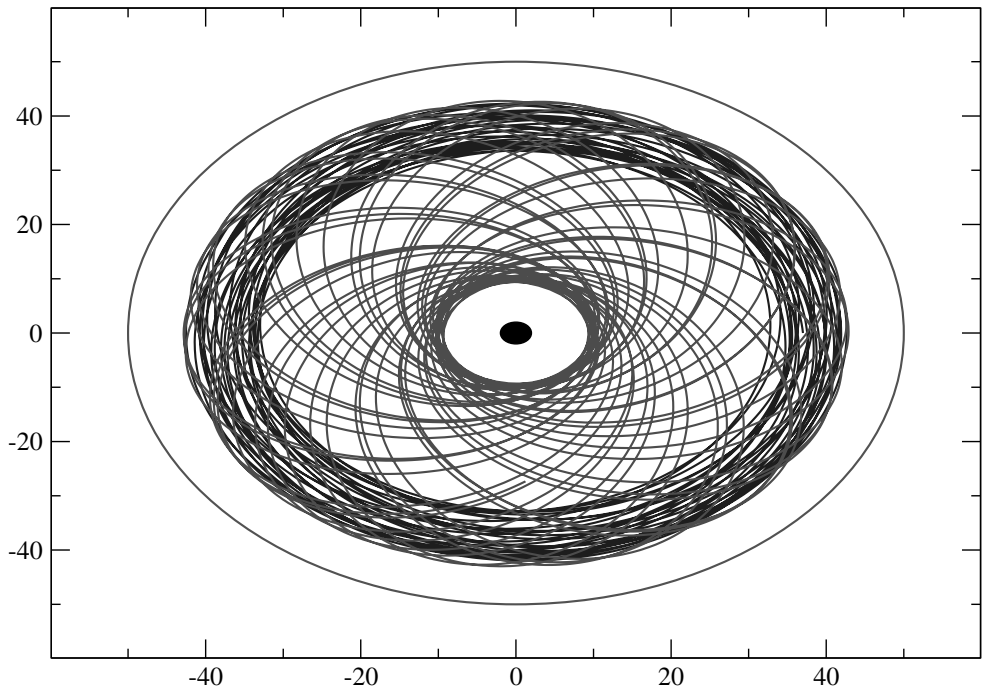}
\caption{A chaotic orbit and a regular one
  between the trajectory of the 
  second body and the black hole are plotted (the orbits are in the $x-y$
  plane). It is clear that both are bounded.}  
\vspace*{1cm}
\label{naocai}
\end{figure}

Some orbits of the test-particle are presented. For the constants presented in
Fig.\ref{escapa}, the unbounded orbits tend to fall into the black hole as we
see in Fig.\ref{cai}. Similar initial conditions but using constants defined
in Fig.\ref{perto} lead to bounded orbits as presented in Fig.\ref{naocai}.

\begin{figure}[p] 
\includegraphics[width=13cm]{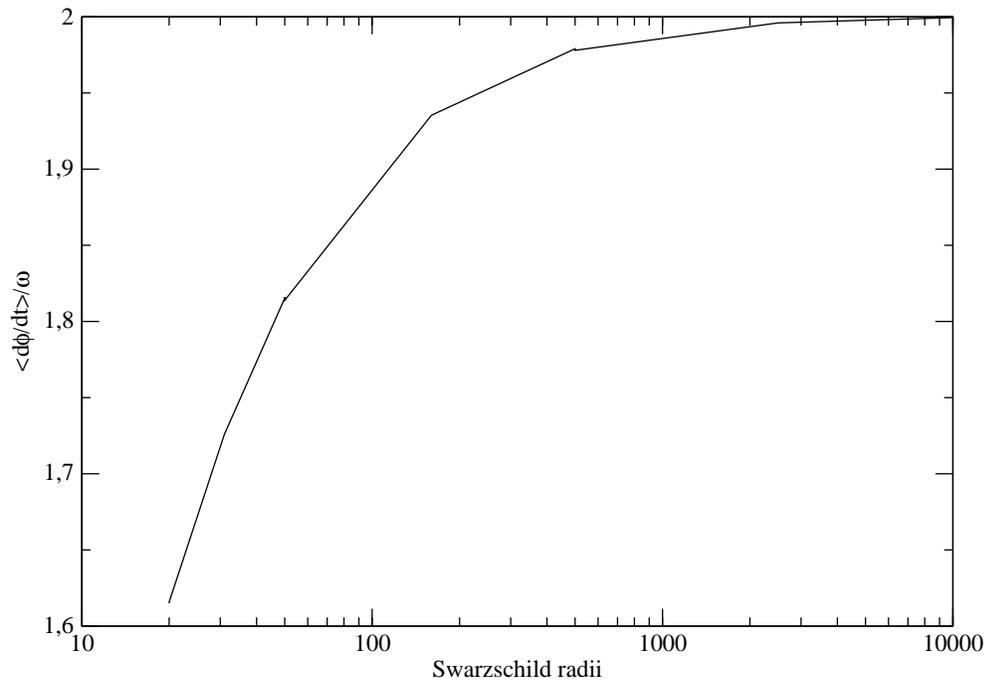}  
\caption{Here we have $\langle d\varphi/dt\rangle/\omega$  obtained for orbits
  that are, in the Newtonian limit, inside  the 2:1 resonant region. For large
  distances to the central mass the ratio of the mean orbital frequency is
  almost 2 whereas near the source (larger relativistic contributions) this
  ratio decreases (indicating loss of commensurability).} 
\label{curva} 
\end{figure}

A remarkable result concerns the commensurability of the stable
orbits. For a typical Newtonian situation (similar to the one presented in
Fig.\ref{longe}) the stable islands usually correspond to regions of
commensurability of periods, i.e., the ratio between the frequency of any
orbit inside this region and the frequency of the second body (that is fixed)
is a rational number. However, when the general relativistic effects increase
this assumption is generally not true although the stable islands are not
easily destroyed as we see in Fig.\ref{perto}. Therefore the
ratio between the period of the test-particle and second body orbits
gradually departs from the rational number obtained in the Newtonian
limit. We shall notice that even close to the black hole,  there is one
specific real number corresponding to each stable island. (The numerical error
when two different orbits in the same island are compared is smaller than
typical fluctuations of the integration routine.) We present a curve of the
ratio between the orbital frequency of the test-particle and the second body
 inside the region corresponding to the Newtonian 2:1 resonance,
 Fig.\ref{curva}. The ordinate is the angular speed of the test-particle
 divided by the orbital frequency of the second body (this value is computed
 after many cycles). The abscissa is the distance between the second body and
 the central source (because it moves circularly). It is very clear that the
 ratio tends to 2 when  $r \rightarrow \infty$ (the Newtonian limit)  and
 decreases as much as we approach the source. This behavior is similar in
 all the stable islands preserved close to the black hole except inside the
 3:2 region where, by construction, the commensurability remains.

\section{Final Remarks}

Although other resonances were studied, the particular case of the circular
restricted three body problem presented in this article  happens to be
the most interesting because large chaotic regions are observed in the
Newtonian case. When we study orbits around the central 2:1 resonance, for
instance, a small chaotic region is noticed for eccentric orbits computed very
close to the central source. Anyway, nothing more remarkable is obtained in
the circular restricted three body problem close to other resonances. 

\begin{figure}[p] 
\includegraphics[width=13cm]{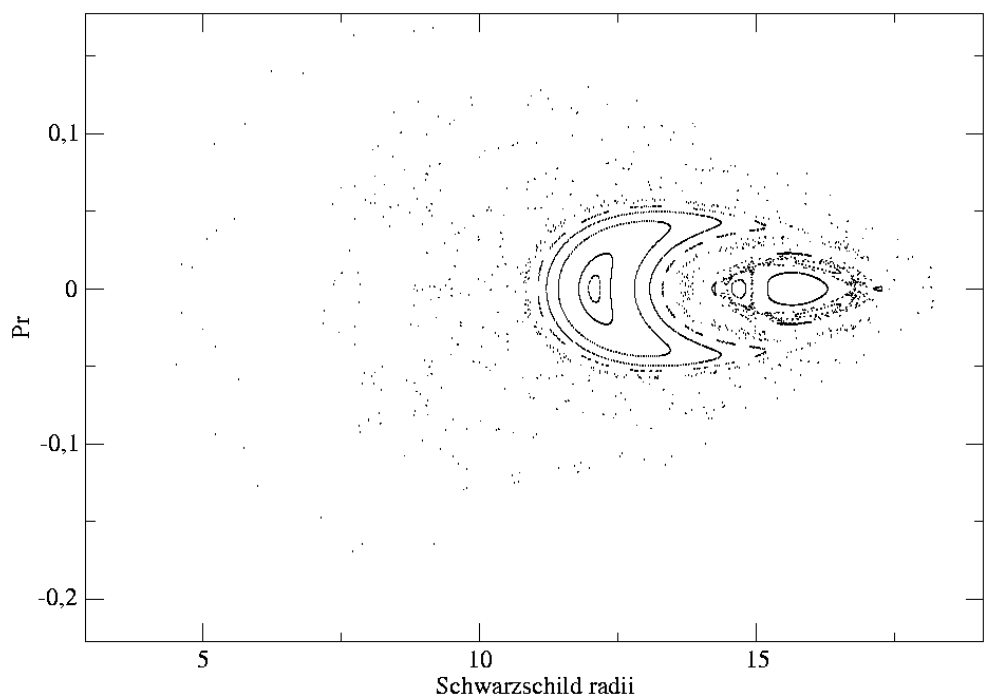}  

\caption{In this very relativistic situation ($K=0.92076$, $r_2=20$) we see
  that the only bounded orbits are the ones that belong to a resonant region.} 
   
\label{escapa} 
\end{figure}

\begin{figure}[p]
\includegraphics[width=13cm]{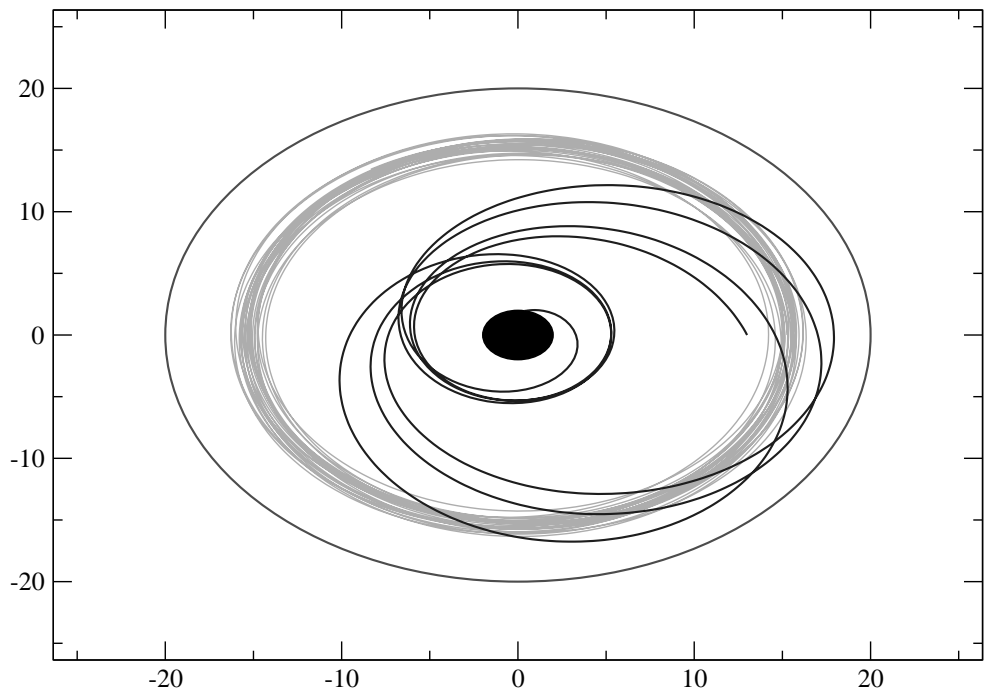}
\caption{Now the orbits are a little bit closer to the black hole. Looking to
  the orbital plane we see that the regular
  orbits remain bounded while the chaotic ones falls into the black hole. This
  situation corresponds to the one presented in the Fig.\ref{escapa}}. 
\label{cai}
\vspace*{0.8cm}
\end{figure}

Despite  only few examples of the restricted three body problem were studied,
I conjecture that the result concerning the 
resonant islands is general. Imagine a situation in which massive objects are
growing by collisions similarly to the  planetary formation
process~\cite{aarseth}.  In this sort of process, lots of particles of dust
remain after some time. From the results here presented it is possible to argue
that most of the  particles were   captured by
the black hole (similar to what is shown in the Fig.\ref{escapa}) because of
the  perturbative influence of  other bodies. On the other hand the objects
formed during this process  should stay in a resonant region (otherwise they
would not be formed). Certainly, a more specific and
complete work must be done in order to verify this speculation.  Since the
technique is based on a Hamiltonian formalism, symplectic  maps should be
used in order to study such complex systems. 

I shall emphasize that the proposed technique 
may be employed (carefully) in different contexts.  The mainly condition is the
predominance of the mean gravitational field associated to the
metric. An interesting application of this method consists in the evolution of
self-interacting systems  in a cosmological background (see for instance the
study of formation of binaries performed by Ioka {\it et al}~\cite{ioka}).
 At last I may say that the aim of this article was mainly to present a new
 approach to study relativistic effects in some astrophysical systems and the
 new results concerning the three body problem although interesting were
 important to test the method rather than presenting a great novelty in
 general relativity.

\ack
I thank to  P. Letelier for the
comments and also  to S. Oliveira, R. Vilela and C. Ghezzi for
carefully reading the paper. I am grateful for the hospitality from EAPS-MIT
and  J. Wisdom for the discussions. This work was supported by FAPESP.

\clearpage

\end{document}